\documentclass[aps,prb,reprint,groupedaddress]{revtex4-2}
\usepackage{float}
\usepackage{mathrsfs}
\usepackage{graphicx,amsfonts,bm,amssymb,amsmath,braket,hyperref}
\hypersetup{colorlinks=true,allcolors=[rgb]{0,0,1}}
\usepackage{mathptmx}
\usepackage{pgfplots}
\usepackage{booktabs}
\pgfplotsset{compat}
\begin{document}

\title{Pairing transitions in a Binary Bose Gas}
\author{Zesheng Shen, Lan Yin}
\affiliation{School of Physics, Peking University, Beijing 100871, China}
\date{\today}
\begin{abstract}
The stable Bardeen-Schrieffer-Cooper (BCS) pairing state of a bosonic system has long been sought theoretically and experimentally. Here we study the BCS state of a binary Bose gas with $s$-wave intra-species repulsions and an inter-species attraction in the mean-field-stable region. We find that above the Bose-Einstein-Condensation (BEC) transtion temperature, there is a phase transtion from the normal state to the BCS state due to inter-species pairing.  When the temperature decreases, another phase transtion from the BCS state to the mixture state with both atomic BEC and inter-species pairs occurs.  As the temperature is further lowered, the mixuture state is taken over by the BEC state.  The phase diagram of this system is presented and experimental implications are discussed.
\end{abstract}
\maketitle
\textit{Introduction}--
The study of superfludity and superconductivity has been a cornerstone of modern condensed matter physics.  Ever since the experimental realization of BEC \cite{Anderson1995,Davis1995,Stringari}, ultracold atoms have provided a new platform for these studies.
By the Feshbach resonance technique, the interaction between atoms system can be tuned, and the BEC-BCS crossover of fermions \cite{crossover1969,Giorgini2008} was experimentally achieved \cite{Greiner2003,Ketterle2003,Jochim2003,Regal2004,Chin2004,Kinast2004,Zwierlein2004,Bourdel2004,Bartenstein2004,Zwierlein2005}.
Compared to the BCS state of fermions \cite{BCS1957}, the pairing state of bosons \cite{Evans1973} was also predicted many years ago, but has never been realized experimentally.
In a single-component Bose gas, it was found theoretically that the BCS pairing state is mechanically unstable with the attractive interaction and the molecular condensation can be stable with the repulsive interaction \cite{mbec1,mbec2,mbec3,mbec4,mbec5,mbec6}.
Experimentally, the strong three-body loss process near the Feshbach resonance has been a major difficulty to create the molecular BEC state \cite{3body1,3body2,3body3,3body4}.
In 2021, a molecular BEC was first experimentally observed in a two-dimensional Bose gas with g-wave closed-channel molecules \cite{cc}. \par

In recent years, the binary Bose gas has attracted a lot of attention, due to the successful experimental realization of quantum droplets \cite{droplet1,droplet2,droplet3}.
In such a system, the inter-species attraction is stronger than the geometric mean of the intra-species repulsion. Although the overall mean-field energy is attractive, the mechanical stability is restored by the Lee-Huang-Yang energy from Gaussian fluctuations \cite{Petrov1,Petrov2}.
In the Bogoliubov theory of the quantum droplet, the phonon excitation energy is imaginary in the long-wavelength limit, implying instablity.
It was later found that the phonon energy is stabilized by higher-order quantum fluctuations \cite{Guqi,Xiong,Zhang}.
In an alternative proposal the ground state of the quantum droplet is predicted to a pairing state rather than the BEC state \cite{huhui1,huhui2}. \par

\begin{figure}[h]
\begin{center}
\includegraphics[width=\linewidth]{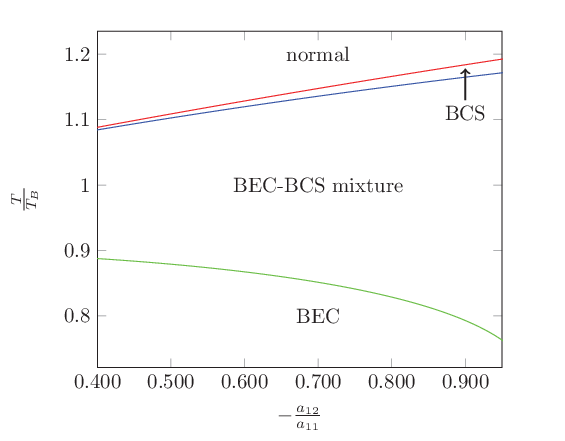}
\caption{Phase diagram of a binary Bose gas for $(na_{11}^3)^{-\frac{1}{3}}=10$, where the horizontal axis is the negative of the ratio of the inter- to intra-species scattering lengths, $n$ is the density of one species, and $T_B=2\pi\hbar^2 n^\frac{2}{3}/[m \zeta^\frac{2}{3}(\frac{3}{2})]$ is the ideal BEC temperature.  Critical temperatures of Normal-BCS, BCS-Mixture and Mixture-BEC transitions are marked by the red, blue and green lines.}
\label{Fig1}
\end{center}
\end{figure}

In this work, we theoretically investigate a dilute binary Bose gas with an inter-species attraction and symmetric intra-species repulsions in the mean-field stable region where the overall mean-field energy is repulsive and dominant over the LHY energy, different from the quantum-droplet case where the mean-field energy is attractive and of the same order of the LHY energy.  We obtain the phase diagram of this system as shown in Fig. \ref{Fig1}.  By studying inter-species pairing self-consistently in mean-field approximation, we find that a stable BCS state exists above the BEC transition temperature and would turn into the normal state as temperature increases.  The gap in the excitation spectrum of the BCS state closes as the temperature decreases to a critical temperature where the phase transition from the BCS state to a mixture state of atomic BEC and inter-species pairs takes place.  The mixture state is taken over by the BEC state at another smaller critical temperature below.  We also discuss how to observe the BCS and mixture states in experiments near the end.
\par
\par
\textit{General Model}--
We consider an uniform binary Bose gas with the Hamiltonian given by $ H=H_0+H_{\text{I}} $.
The single-particle part is given by
\begin{equation}
	H_0 = \sum_{\mathbf{k},i} \epsilon_{\mathbf{k}}  c_{i\mathbf{k}}^{\dagger} c_{i\mathbf{k}},
\end{equation}
and the $s$-wave interaction term is given by
\begin{align}
	H_{\text{I}} &= \sum_{\mathbf{k}_1\mathbf{k}_2\mathbf{k}_3,i} \frac{g_{ii}}{2V}c_{i\mathbf{k}_1}^{\dagger}c_{i\mathbf{k}_2}^{\dagger}c_{i\mathbf{k}_3}c_{i\mathbf{k}_1+\mathbf{k}_2-\mathbf{k}_3} \nonumber \\
	&+ \sum_{\mathbf{k}_1\mathbf{k}_2\mathbf{k}_3}' \frac{g_{12}}{V}c_{1\mathbf{k}_1}^{\dagger}c_{2\mathbf{k}_2}^{\dagger}c_{2\mathbf{k}_3}c_{1\mathbf{k}_1+\mathbf{k}_2-\mathbf{k}_3},
\end{align}
where $V$ is the volume, $\epsilon_{\mathbf{k}}=\hbar^2\mathbf{k}^2/2m$, $c_{i\mathbf{k}}$ is the boson annihilation operator of the $i$-th component, $i=1,2$, $g_{ij} = 4\pi\hbar^2 a_{ij}/m$ is the coupling constant between $i$- and $j$-th components, and $a_{ij}$ is the scattering length.  In the following, we focus on the case with the inter-species attraction $a_{12}<0$ and symmetric intra-species repulsions $a_{11}=a_{22}>0$ in the mean-field stable region  $|a_{12}|<a_{11}$.
\par

\textit{BCS pairing state}--
We consider pairing due to the attractive inter-species interaction and define the pairing order parameter as
\begin{equation}
\Delta = \frac{g_{12}}{V}\sum_{\mathbf{k}} \langle c_{1\mathbf{k}}c_{2-\mathbf{k}}\rangle,
\end{equation}
where the term with $\mathbf{k}=0$ is negligible in the thermodynamical limit in the absence of a BEC and should be treated separately in the presence of a BEC.  We first study the pure BCS state and set $\Delta$ to be a negative real number without losing generality. In the mean-field approximation which include the pairing and Hartree-Fock energies, the Hamiltionian in the grand-canonical ensemble is given by
\begin{equation} \label{BCS}
	H_{B}-\mu N= \sum_\mathbf{k}{ \begin{pmatrix}  c_{1\mathbf{k}}^{\dagger},c_{2-\mathbf{k}} \end{pmatrix}}
	{ \begin{pmatrix} \epsilon_\mathbf{k}-\mu^{\prime}&\Delta \\ \Delta&\epsilon_\mathbf{k}-\mu^{\prime} \end{pmatrix}} { \begin{pmatrix}  c_{1\mathbf{k}} \\ c_{2-\mathbf{k}}^{\dagger} \end{pmatrix}} +E_B,
\end{equation}
where
$$E_B=-\sum_\mathbf{k} \left(\epsilon_\mathbf{k}-\mu^{\prime}\right) - \frac{\Delta^2V}{g_{12}} - \left(2g_{11}+g_{12}\right)n^2 V,$$
$\mu$ is the chemical potential, $\mu^{\prime} = \mu - \left(2g_{11}+g_{12}\right)n$ is the shifted chemical potential excluding the Hartree-Fock energy, and $N$ is the total number operator.

This mean-field Hamiltonian can be diagonalized by the Bogoliubov transformation with the quasi-particle excitation energy given by
\begin{equation}
	E_\mathbf{k} = \sqrt {\left( \epsilon_\mathbf{k}-\mu^{\prime} \right)^2 - \Delta^2}, \label{E_k of bcs}
\end{equation}
showing that the minimum excitation energy has a gap given by $E_0=\sqrt{\mu'^2-\Delta^2}$.
The shifted chemical potential $\mu'$ and the pairing order parameter $\Delta$ can be self-consistently solved from the following number and gap equations,
\begin{subequations}
	\begin{align}\label{eq of BCS_1}
		n & = \int \frac{d^3 \mathbf{k}}{2\left(2\pi\right)^3}\left( \frac{\epsilon_\mathbf{k}-\mu^{\prime}}{E_\mathbf{k} \tanh\frac{\beta E_\mathbf{k}}{2}} - 1\right), \\
		-\frac{m}{4\pi\hbar^2 a_{12}} & = \int \frac{d^3 \mathbf{k}}{2\left(2\pi\right)^3}\left(\frac{1}{E_\mathbf{k} \tanh\frac{\beta E_\mathbf{k}}{2}} - \frac{1}{\epsilon_\mathbf{k}}\right),\label{eq of BCS_2}
	\end{align}
\end{subequations}
where we have used the renormalization relation of the $s$-wave coupling constant
\begin{equation}
	\frac{1}{g_{12}}=\frac{m}{4\pi\hbar^2 a_{12}}-\frac{1}{2V}\sum_\mathbf{k}\frac{1}{\epsilon_\mathbf{k}}.
\end{equation}

\begin{figure}[h]
	\begin{center}
		\includegraphics[width=0.8\columnwidth]{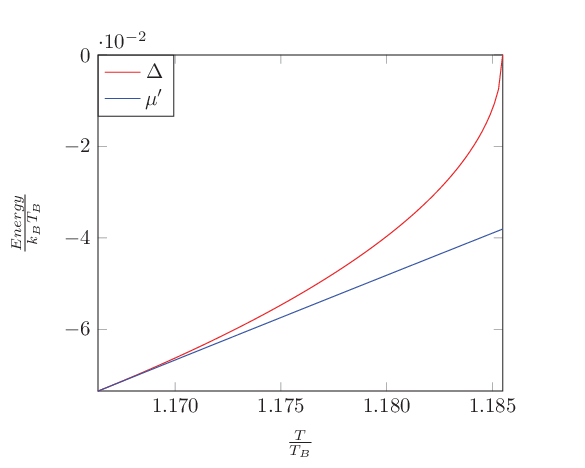}
		\caption{Pairing order parameter $\Delta$ and shifted chemical potential $\mu^\prime$ versus temperature, numerically calculated at $(na_{11}^3)^{-\frac{1}{3}}=10$ and $(na_{12}^3)^{-\frac{1}{3}}=-11$.  As the temperature decreases, both $\Delta$ and $\mu^{\prime}$ decrease.  The transtion from the normal state to the BCS state occurs at $T_{C1} \approx 1.186 T_B $, and the transition from the BCS state to the mixture state takes place at $T_{C2} \approx 1.166 T_B$.}
		\label{Fig2}
	\end{center}
\end{figure}

The above equations are numerically solved, and the BCS state generally exists between two critical temperatures, $T_{C2}\leq T \leq T_{C1}$, as shown in Fig. \ref{Fig2}.  At the first critical temperature $T_{C1}$, the pairing order parameter $\Delta$ vanishes, and the phase transition from the normal state to the BCS state takes place.  The critical temperature $T_{C1}$  can be obtained from the following $T_C$-equation
\begin{equation}
	-\frac{m}{4\pi\hbar^2 a_{12}}  = \int \frac{d^3 \mathbf{k}}{2\left(2\pi\right)^3}\left(\frac{1}{\epsilon_\mathbf{k}-\mu'} \coth\frac{\epsilon_\mathbf{k}-\mu'}{2k_BT} - \frac{1}{\epsilon_\mathbf{k}}\right)\label{eq of Tc_1}.
\end{equation}
The r.h.s. of Eq. (\ref{eq of Tc_1}) has an infrared divergence at $\mu'=0$ when the temperature $T$ reaches the ideal BEC temperature, $T_B=2\pi\hbar^2 n^\frac{2}{3}/[m \zeta^\frac{2}{3}(\frac{3}{2})]$, showing that Eq. (\ref{eq of Tc_1}) always has a solution $T_{C1}>T_B$ no matter how weak the inter-species interaction is.  Thus starting from the normal state, as the temperature decreases, the system always first enters the BCS pairing state before reaching the BEC state.  As shown in Fig. \ref{Fig2}, at $(na_{11}^3)^{-\frac{1}{3}}=10$ and $(na_{12}^3)^{-\frac{1}{3}}=-11$, the critical temperature $T_{C1}$ is about $1.186T_B$.

In this BCS paring state, as the temperature further decreases, both $|\mu'|$ and $|\Delta|$ increase, but the energy gap $E_0$ drecreases, as shown in Fig. \ref{Fig2}.  At the second critical temperature $T_{C2}$, $\mu'=\Delta$, the gap in the excitation energy vanishes, $E_0$=0, and the system is likely to go into a mixture state of pairs and BEC atoms, which is explored in the latter part of this work.
It can be shown from Eq. (\ref{eq of BCS_1}) and (\ref{eq of BCS_2}) that the critical temperature $T_{C2}$ is also always bigger than $T_B$.
In Fig. \ref{Fig2}, $T_{C2}$ is about $1.166 T_B$.
\par

In the dilute region, both energies, $|\mu'|$ and $|\Delta|$, are much less than $k_BT_B$, as shown in Fig. \ref{Fig2}.  The pair function defined by
$$	\chi(\mathbf{r})=\frac{1}{V} \sum_{\mathbf{k}} \langle c_{1\mathbf{k}}c_{2-\mathbf{k}}\rangle e^{i\mathbf{k}\cdot\mathbf{r}},$$
can be obtained analytically at large distance $r \gg n^{-1/3}$,
\begin{equation}
	\chi(\mathbf{r}) \approx \frac{m k_B T}{4 \pi \hbar^2 r}(e^{-r/\xi}-e^{-r/{\xi'}}),
\end{equation}
where the two characteristic lengths are given by $\xi=(2m |\mu'-\Delta|/\hbar^2)^{-1/2}$ and $\xi'=(2m |\mu'+\Delta|/\hbar^2)^{-1/2}$.  In the dilute region the pair size $\xi$ is much larger than the interparticle distance, $\xi \gg n^{-1/3}$, indicating that there are big spatial overlaps of the pairs.  The pairing state is clearly in the BCS limit, opposite to the molcular BEC limit.

\textit{Mixture state}--
At the lower critical temperature $T_{C2}$, the energy gap preventing the BEC formation vanishes, $E_0=0$.  Here we explore the possibility of the mixture state with atom BEC and phase-coherent inter-species pairs below $T_{C2}$.  In the mixture state, two order parameters coexist, the BEC wavefunctions $\psi_{0}=\langle c_{i0} \rangle /\sqrt{V}$ and the pairing order parameter $\Delta= g_{12}\sum_{\mathbf{k}\neq 0} \langle c_{1\mathbf{k}}c_{2-\mathbf{k}}\rangle/V$.  We consider the case that the BEC and pairs are phase coherent, and assume $\psi_0>0$ and $\Delta <0$ without losing generality.  By including the Hartree-Fock and pairing energies, we obtain the mean-field Hamiltonian in the grand-canonical ensemble given by
\begin{align}\label{mix}
H_M-\mu N &= E_M +  \sum_{\mathbf{k} \neq 0} \Bigl[\left(\epsilon_\mathbf{k}-\mu^{\prime}\right)\left(c_{1\mathbf{k}}^{\dagger} c_{1\mathbf{k}} + c_{2\mathbf{k}}^{\dagger} c_{2\mathbf{k}}\right) \nonumber\\
 + &\frac{g_{11}n_0}{2}\left(c_{1\mathbf{k}}^{\dagger} c_{1-\mathbf{k}}^{\dagger} + c_{2\mathbf{k}}^{\dagger} c_{2-\mathbf{k}}^{\dagger} + h.c.\right) \nonumber\\
 + & g_{12}n_0\left(c_{1\mathbf{k}}^{\dagger}c_{2\mathbf{k}}+ h.c.\right) +  (\Delta+g_{12}n_0)\left(c_{1\mathbf{k}}c_{2-\mathbf{k}}+ h.c.\right)  \Bigr],
\end{align}
\vspace{-\baselineskip}
where $$\frac{E_M}{V}=g_{11}\left(n_0^2-2\tilde{n}^2\right) + g_{12}[n_0^2-\tilde{n}^2 - \left(\frac{\Delta}{g_{12}}\right)^2]-2\mu n_0,$$ $n_0=\psi_0^2$ is the condensate density of one species, $\tilde{n}=n-n_0$ is the atom density outside the condensate of one species, and $\mu^{\prime} = \mu - (2g_{11}+g_{12})n$ as defined before.  There are important differences between Eq. (\ref{mix}) and the Hamiltonian of the pairing state proposed for quantum droplets \cite{huhui1,huhui2}, i. e. in Eq. (\ref{mix}) the Hartree-Fock energy from the inter-species interaction is included and the mean-field contribution from non-condensed atoms is taken into account as in Popov's approximation \cite{popov}.  Eq. (\ref{mix}) is capable to desribe all the three broken-symmetry states, i. e. BEC, BCS and mixture states.  When $\Delta=0$, Eq. (\ref{mix}) recovers the Bogoliubov Hamiltonian of the BEC state; when $n_0=0$, Eq. (\ref{mix}) recovers the Hamiltonian of the BCS state, Eq. (\ref{BCS}).

After the canonical transformation $\alpha_\mathbf{k}^{\dagger} = \frac{1}{\sqrt{2}}\left(c_{1\mathbf{k}}^{\dagger}+c_{2\mathbf{k}}^{\dagger}\right)$, $\beta_\mathbf{k}^{\dagger} = \frac{1}{\sqrt{2}}\left(c_{1\mathbf{k}}^{\dagger}-c_{2\mathbf{k}}^{\dagger}\right)$,  Eq. (\ref{mix}) becomes
\begin{align}\label{mixp}
H_M-\mu N  & =\frac{1}{2} \sum_{\mathbf{k} \neq 0} \Bigl[\left(\epsilon_\mathbf{k}-\mu^{\prime}+g_{12}n_0\right)\left(\alpha_\mathbf{k}^{\dagger} \alpha_\mathbf{k}  +\alpha_{-\mathbf{k}}^{\dagger} \alpha_{-\mathbf{k}}\right)\nonumber\\
+&\left((g_{11}+g_{12})n_0+\Delta\right)\left(\alpha_\mathbf{k}^{\dagger}\alpha_{-\mathbf{k}}^{\dagger}+h.c.\right)\Bigr] \nonumber\\
+&\frac{1}{2} \sum_{\mathbf{k} \neq 0} \Bigl[\left(\epsilon_\mathbf{k}-\mu^{\prime}-g_{12}n_0\right)\left(\beta_\mathbf{k}^{\dagger} \beta_\mathbf{k}  +\beta_{-\mathbf{k}}^{\dagger} \beta_{-\mathbf{k}}\right)\nonumber\\
+&\left((g_{11}-g_{12})n_0-\Delta\right)\left(\beta_\mathbf{k}^{\dagger}\beta_{-\mathbf{k}}^{\dagger}+h.c.\right)\Bigr] +E_M.
\end{align}
The mean-field Hamiltonian in Eq. (\ref{mixp}) can be diagonalized by Bogoliubov transformation and two excitation braches are obtained.  The density-excitation energy is given by
\begin{equation}
	E_{\mathbf{k}+} = \sqrt {\left(\epsilon_\mathbf{k}-\mu'+g_{12}n_0 \right)^2-[(g_{11}+g_{12})n_0+\Delta]^2},
\end{equation}
and the spin-excitation energy is given by
\begin{equation}
	E_{\mathbf{k}-} = \sqrt {\left(\epsilon_\mathbf{k}-\mu'-g_{12}n_0 \right)^2-[(g_{11}-g_{12})n_0-\Delta]^2}.
\end{equation}
From the mean-field thermodynamical potential
\begin{align}
	\Omega_0 &=\langle H_I-\mu N \rangle \nonumber \\
            &=V \{g_{11}[n_0^2+\tilde{n}(4n_0+2\tilde{n})]+g_{12}[(n_0+\tilde{n})^2+\frac{\Delta}{g_{12}}^2] \notag \\
            &-2\mu (n_0+\tilde{n}) \},
\end{align}
the chemical potential $\mu$ can be obtained by the minimization condition $\frac{\partial\Omega_0}{\partial n_0}|_{\tilde{n},\Delta}=0$,
\begin{equation} \label{mcp}
	\mu=(g_{11}+g_{22})n_0+(2g_{11}+g_{12})\tilde{n}+\Delta.
\end{equation}
The excitation energies are thus given by
\begin{align}
    E_{\mathbf{k}+} &= \sqrt {\left(\epsilon_\mathbf{k}-2\Delta\right)[\epsilon_\mathbf{k}+2(g_{11}+g_{12})n_0]^2} \label{E_k of mix_1},\\
    E_{\mathbf{k}-}&= \sqrt {\epsilon_\mathbf{k}[\epsilon_\mathbf{k}+2(g_{11}-g_{12})n_0-2\Delta]^2} \label{E_k of mix_2},
\end{align}
showing that the density excitation is gapped as in BCS state while the spin excitation is now gappless as found in Ref \cite{huhui1,huhui2}, but the detailed spectrum are different.  Especially the density-excitation energy $E_{\mathbf{k}+}$ becomes unstable at the mean-field-unstable point, $g_{11}+g_{12}=0$.  This difference is due to the inclusion of the Hartree-Fock term from the inter-species interaction to the mean-field Hamiltonian in our treatment as mentioned above.

The condensation density $n_0$ and the pairing order parameter $\Delta$ can be further obtained from the self-consistent condition
\begin{widetext}
	\begin{subequations}
		\begin{align}
			&\frac{\Delta}{4\pi\hbar^2 a_{12}/m} = \int\frac{d^3\mathbf{k}}{2\left(2\pi\right)^3}\Bigl[\left(\sum_\pm \pm\frac{g_{11}n_0\pm(\Delta+g_{12}n_0)}{2E_{\mathbf{k}\pm}}\left(2f_{\mathbf{k}\pm}+1\right)\right)+\frac{\Delta+g_{12}n_0}{\epsilon_\mathbf{k}}\Bigr] \label{eq of mix_1}, \\
			& n_0 = n  -\int\frac{d^3\mathbf{k}}{2\left(2\pi\right)^3}\sum_\pm \Bigl[\frac{\epsilon_\mathbf{k}+\left(g_{11} \pm g_{12}\right)n_0-\Delta}{2E_{\mathbf{k}\pm}}\left(f_{\mathbf{k}\pm}+\frac{1}{2}\right)-\frac{1}{2}\Bigr], \label{eq of mix_2}
		\end{align}
	\end{subequations}
where $\sum_\pm$ stands for summation over two spectrums, $f_{\mathbf{k}\pm}=(e^{\beta E_{\mathbf{k}\pm}}-1)^{-1}$  are Bose distribution functions of quasi-particles.
\end{widetext}
\par

\begin{figure}[h]
    \centering
    \includegraphics[width=0.8\columnwidth]{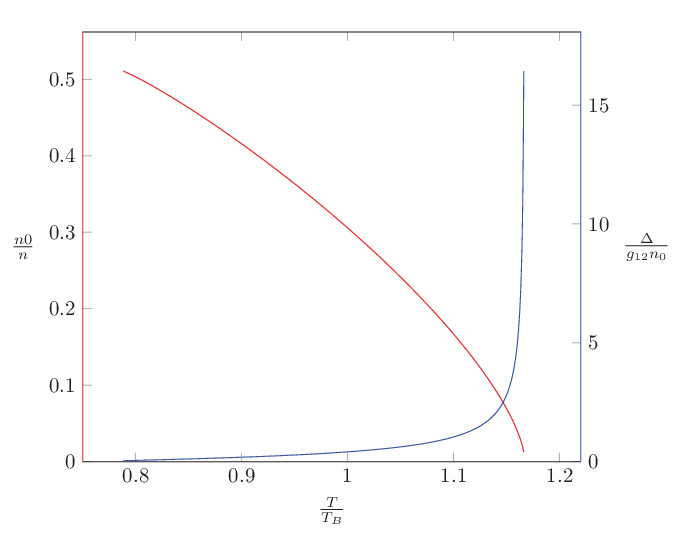}
    \caption{The BEC fraction $n_0/n$ and the pairing fraction $\Delta/(g_{12}n_0)$ versus temperature in the mixture state, numerically calculated at $(na_{11}^3)^{-\frac{1}{3}}=10$ and $(na_{12}^3)^{-\frac{1}{3}}=-11$. As the temperature decreases, the BEC fraction increases and the pairing fraction decreases, indicating that the system is turning into a BEC state.}
    \label{Fig3}
\end{figure}

\begin{figure}[h]
	\centering
    \includegraphics[width=\linewidth]{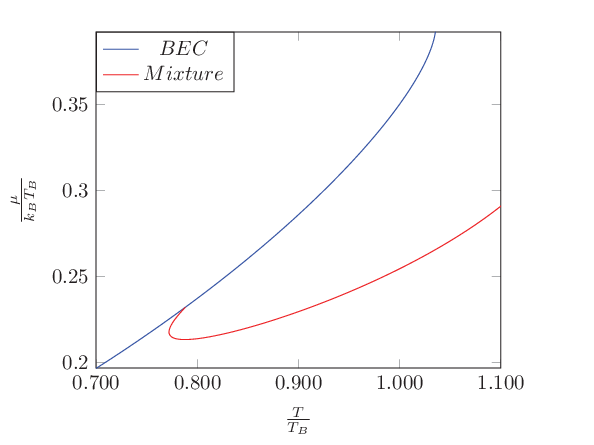}
    \caption{Chemical potential $\mu$ in the mixture state and BEC state versus temperature for $(na_{11}^3)^{-\frac{1}{3}}=10$ and $(n a^3_{12})^{-\frac{1}{3}}=-11$.  A first-order phase transition from the BEC state to the mixture state occurs at about $0.788T_B$.}
    \label{Fig4}
\end{figure}

The numerical solutions of these two equations at $(na_{11}^3)^{-\frac{1}{3}}=10$ and $(na_{12}^3)^{-\frac{1}{3}}=-11$ are shown in Fig. \ref{Fig3}. Below the critical temperatures $T_{C2}$, both the condensation fraction $n_0/n$ and the pairing fraction $\Delta/(g_{12}n_0)$ decrease with the temperature, showing that the mixture state evolves towards the BEC state. At low temperatures, the pairing fraction is almost negligible and a transition into the BEC state is likely to occur. By comparing the chemical potential of the BEC state \cite{popov} with that of the mixture state from Eq. (\ref{mcp}) in Fig. \ref{Fig4}, we find that a first-order phase transition occurs at about $0.788T_B$ for the same parameters. Fig. \ref{Fig4} also shows a cusp in the chemical potential of the mixture state near the BEC state, which is probably a mean-field artifact and may be corrected by the higher-order fluctuation effect.

\textit{Discussion and conclusion}--
Quantum droplets have been experimentally realized in homonuclear $^{39}\mathrm{K}$ binary gases \cite{39K1,39K2,39K3} in the the mean-field-unstable region with $-a_{12}>\sqrt{a_{11}a_{22}}$ where the LHY energy is of the same order of the mean-field energy.  Our phase diagram is in the mean-field-stable region and can be tested in the same experimental setup by tunning the intra-species interactions symmetric $a_{11}=a_{22}$ in the region $0<-a_{12}<\sqrt{a_{11}a_{22}}$ with Feshbach resonance.  For the asymmetric case, our results about the BCS state can still be tested with the detuning energy $\delta=2(g_{11}-g_{22}) n$ in the 2nd component, so that both components are equally populated, equivalent to the symmetric case, while the mixture state is more complicated.

The BEC, BCS, and mixture states can be distinguished in their excitation spectrum.  In the BEC state, both spin and density excitations are gapless; in the BCS state, both excitations are gapful; in the mixture state, the spin excitation is gapless, while the density excitation is gapped.  These three states also have differences in their topological excitations.  In a vortex of the BEC state, the angular momentum per atom in the condensate is $\hbar$; in a vortex of the BCS state, the angular momentum per pairing atom is reduced to half.

The mixture state in the mean-field stable region is rather similar to the pairing state in the mean-field unstable region proposed for the quantum droplet \cite{huhui1}.  Both states contain atomic condensation and interspecies pairs, but subject to slightly different treatments.  Here we have included all the Hartree-Fock energies, especially the Hartree-Fock energy from the inter-species interaction which was omitted in Ref. \cite{huhui1,huhui2}.   The mixture state exists only at finite temperatures and is taken over by the BEC state at low temperatures.  In contrast, the pairing state of the quantum droplet exists at zero temperature and is destabilized at a finite temperature \cite{huhui3}.  The boson pairing state is also related to the fermionic quartet \cite{quartet}.

In conclusion, we obtain the phase diagram of a symmetric Bose gas with the attractive inter-species interaction in the mean-field stable region and three pairing-related phase transitions are identified.  As the temperature decreases, the system first turns from the normal state to the BCS pairing state.  As the temperature continues to decrease, a phase transition from the BCS state to a mixture of BEC and pairs takes place.  At a temperature further below, a first-order phase transition from the mixture state to the BEC state occurs.  Our results may be tested in current experimental setups.

\begin{acknowledgments}
	We would like to thank T.-L. Ho, Z.-Q. Yu, and Q. Gu for helpful discussions.
\end{acknowledgments}

\end{document}